\documentclass[a4paper]{jpconf}

\usepackage{graphicx}
\usepackage{relsize}

\def\ra     {\!\rightarrow\!}
\def\cp     {$CP$}
\def\infb     {fb$^{-1}$}

\def\babar{\mbox{\slshape B\kern-0.1em{\smaller A}\kern-0.1em
    B\kern-0.1em{\smaller A\kern-0.2em R}}}

\begin{document}

\title{
Recent results from the Belle experiment \\
\vskip-2.2in
\hskip5.2in \normalsize{\rm UCHEP-16-07} 
\vskip2.0in
}

\author{Z King and A J Schwartz \\
(on behalf of the Belle Collaboration)}

\address{
Physics Department, University of Cincinnati, P.O. Box 210011,
Cincinnati, Ohio 45221 USA }

\ead{kingze@mail.uc.edu, alan.j.schwartz@uc.edu}

\begin{abstract}
We review recent results from the Belle experiment, which took 
data at the KEKB asymmetric-energy $e^+ e^-$ collider in Japan. 
The experiment recorded about 1000~fb$^{-1}$ of data running 
mainly at the $\Upsilon(4S)$ and $\Upsilon(5S)$ resonances. 
The results presented here are obtained from the full data set.
\end{abstract}

\section{Introduction}

The Belle experiment successfully operated for more than a decade until
2010 at the asymmetric-energy $e^+e^-$ collider KEKB~\cite{KEKB}.
The experiment took data at center-of-mass energies corresponding to
several $\Upsilon(n$S) resonances; the total data sample recorded 
exceeds~$1\,\mathrm{ab}^{-1}$. Here we present recent results 
based on the full data sample.

The Belle detector is a large-solid-angle magnetic spectrometer that
consists of a silicon vertex detector (SVD), a 50-layer central drift
chamber (CDC), an array of aerogel threshold Cherenkov counters (ACC), a
barrel-like arrangement of time-of-flight scintillation counters (TOF),
and an electromagnetic calorimeter comprised of CsI(Tl) crystals (ECL)
located inside a super-conducting solenoid coil that provides a 1.5~T
magnetic field. An iron flux-return located outside of the coil is
instrumented to detect $K^0_L$ mesons and to identify muons (KLM). The
detector is described in detail elsewhere~\cite{BelleDetector}.


\section{Search for the Decay $B^0 \rightarrow \phi \gamma$}
\label{sec:rho_gamma}

In the Standard Model (SM), the decay
$B^{0} \ra \phi\gamma$ proceeds through
electroweak and gluonic $b\ra d$ penguin annihilation processes.
These amplitudes are proportional
to the small Cabibbo-Kobayashi-Maskawa~\cite{CKM} matrix element
$V_{td}$ and thus are highly suppressed.
The branching fraction has been estimated based on naive QCD 
factorization~\cite{QCD} and perturbative QCD~\cite{pQCD}
and found to be in the range $10^{-12}$ to $10^{-11}$.
However, the internal loop can also be mediated by non-SM
particles such as a charged Higgs boson or supersymmetric squarks,
and thus the decay is sensitive to new physics (NP). It is estimated
that such NP could enhance the branching fraction to the level of 
$10^{-9}$ to $10^{-8}$~\cite{QCD}.
Previously, no evidence for this decay has been found. Here
we describe a recent Belle search for this decay 
with much higher sensitivity than previously that uses
the full data set of 711~\infb\ recorded on the $\Upsilon$(4S) 
resonance~\cite{King:2016cxv}.

Candidate $\phi$ mesons are reconstructed via $\phi \ra K^+K^-$
decays. The $K^+K^-$
invariant mass is required to be in the range $[1.000, 1.039]$~GeV/$c^2$,
which corresponds to $4.5\sigma$ in resolution around the $\phi$ mass.
Candidate $B$ mesons are identified using a modified beam-energy-constrained
mass $M_{\rm bc} = \sqrt{E^2_{\rm beam} - |\vec{p}^{}_B|c^2}/c^2$, and the energy
difference $\Delta E = E_B - E_{\rm beam}$, where $E_{\rm beam}$ is the beam
energy and $\vec{p}^{}_B$ and $E_B$ are the momentum and energy, respectively,
of the $B^0$ candidate. All quantities are evaluated in the center-of-mass 
frame. To improve the $M_{\rm bc}$ resolution, the momentum $\vec{p}^{}_B$ 
is calculated as
$\vec{p}^{}_{\phi} + 
(\vec{p}^{}_\gamma/|p^{}_{\gamma}|)\sqrt{(E_{\rm beam} - E_{\phi})^{2}}/c$,
where $\vec{p}^{}_{\gamma}$ is the photon momentum, and $E_{\phi}$ and 
$\vec{p}^{}_\phi$ are the energy and momentum, respectively,
of the $\phi$ candidate. We require that events satisfy 
$5.25~{\rm GeV}/c^2 < M_{\rm bc} < 5.29~{\rm GeV}/c^2$ and 
$-0.30~{\rm GeV}<\Delta E < 0.15~{\rm GeV}$. The signal
yield is calculated in a smaller ``signal region''  
$5.27~{\rm GeV}/c^2 < M_{\rm bc} < 5.29~{\rm GeV}/c^2$
and $-0.20~{\rm GeV}<\Delta E < 0.10~{\rm GeV}$.

Charmless hadronic decays suffer from large backgrounds arising
from continuum $e^+e^- \ra q\bar{q}\; (q = u, d, s, c)$
production. To suppress this background, we use a multivariate
analyzer based on a neural network (NN)~\cite{Feindt:2006pm}.
The NN generates an output variable $C_{\rm NN}$, which ranges from
$-1$ for background-like events to $+1$ for signal-like events.
We make a loose requirement $C_{\rm NN} > 0.3$
and then translate $C_{\rm NN}$ to $C'_{\rm NN}$, defined as
$C'_{\rm NN} = 
\ln\left[ (C_{\rm NN} - C_{\rm min})/(C_{\rm max} - C_{\rm NN})\right]$
where $C_{\rm min}$ = 0.3 and $C_{\rm max}$ = 1.0. 
The variable $C'_{\rm NN}$ is well-described by a sum of Gaussian 
functions. After the above selections, 961 events remain.
The remaining background consists of continuum events and rare
charmless $b$-decay processes.

We obtain the signal yield using an unbinned extended maximum
likelihood fit to the observables $M_{\rm bc}$, $\Delta E$, $C'_{\rm NN}$,
and $\cos\theta_\phi$. The helicity angle $\theta_\phi$ is the angle
between the $K^+$ momentum and the opposite of the $B$ flight direction
in the $\phi$ rest frame. 
The resulting branching fraction is calculated as
\begin{eqnarray}
\mathcal{B}\left(B^0\ra \phi \gamma \right) & = & 
\frac{N_{\rm sig}}
{N_{B\overline{B}} \cdot \varepsilon \cdot \mathcal{B}(\phi \ra K^+K^-)}
\label{eqn:br}
\end{eqnarray}
where $N_{\rm sig} = 3.4\,^{+4.6}_{-3.8}$ is the 
signal yield in the signal region,
$N_{B\overline{B}} = (772 \pm 11) \times 10^6$ is the number
of $B\overline{B}$ events, 
$\varepsilon = 0.296\pm 0.001$ is the signal efficiency 
as calculated from Monte Carlo (MC) simulation, and 
$\mathcal{B}(\phi \ra K^+K^-) = (48.9 \pm 0.5)\%$~\cite{PDG}.
We find no evidence for this decay and set an upper limit
on the branching fraction of 
$\mathcal{B}(B^0 \ra \phi \gamma) < 1.0 \times 10^{-7}$ 
at 90\% confidence level (C.L.).
This limit is almost an order of magnitude lower than 
the previous most stringent result~\cite{babar_phig}.


\section{Angular Analysis of $B^0 \ra K^*(892)^0 \ell^+\ell^-$}

The rare decay $B^0 \ra K^*(892)^0 \ell^+\ell^-$ involves the quark
transition $b\to s \ell^+ \ell^-$, a flavor changing neutral current
that is forbidden at tree level in the SM.
Higher order SM processes such as penguin or $W^+W^-$ box diagrams
allow for such transitions, leading to branching fractions below $10^{-6}$.
Various extensions to the SM predict NP amplitudes that can
interfere with the SM amplitudes and lead to enhanced or 
suppressed branching fractions and modified angular distributions. 

Belle performed an angular analysis using the 
decay modes $B^0 \ra K^*(892)^0 e^+ e^-$ and 
$B^0 \ra K^*(892)^0 \mu^+ \mu^-$~\cite{Abdesselam:2016llu}.
$K^{*0}$ candidates are reconstructed in the channel 
$K^{*0}\to K^+\pi^-$~\cite{charge-conjugates}.
Large combinatoric background is suppressed by 
applying requirements on kinematic variables.
Large backgrounds also arise from charmonium decays 
$B\to K^{(*)} J/\psi$ and $B\to K^{(*)} \psi(2S)$, in which 
$\psi\rightarrow\ell^+\ell^-$. To maximize signal efficiency 
and purity, neural networks are applied sequentially beginning
from the end of the decay chain.
Signal and background yields are determined from an unbinned 
extended maximum likelihood fit to $M_{\rm bc}$ in bins of $q^2$.
The bin ranges used and the resulting fitted yields are 
listed in Table \ref{tab:kll_yields}. In total, $117.6\pm12.4$ signal 
candidates are obtained for $B^0 \ra K^*(892)^0\mu^+\mu^-$, and 
$69.4\pm 12.0$ candidates for $B^0 \ra K^*(892)^0 e^+ e^-$. 

\begin{table}
\caption{\label{tab:kll_yields} 
Fitted yields in bins of $q^2$ for signal ($N_{\rm sig}$) 
and background ($N_{\rm bkg}$), for electron and muon 
channels combined.}
\begin{center}
\begin{tabular}{cccc}
\br
Bin & $q^2$ range in ${\rm GeV}^2/c^4$ &  $N_{\rm sig}$ &  $N_{\rm bkg}$\\ 
\mr
0 & $1.00 - 6.00$  &  $49.5\pm 8.4$  &  $30.3\pm 5.5$  \\
1 &$0.10 - 4.00$  &  $30.9\pm 7.4$  &  $26.4\pm 5.1$   \\
2 &$4.00 - 8.00$  &  $49.8\pm 9.3$  &  $35.6\pm 6.0$   \\
3 &$10.09 - 12.90$  &  $39.6\pm 8.0$  &  $19.3\pm 4.4$ \\
4 &$14.18 - 19.00$  &  $56.5\pm 8.7$  &  $16.0\pm 4.0$ \\
\br
\end{tabular}
\end{center}
\end{table}

We subsequently perform the angular analysis.
The decay is kinematically described  by three angles
($\theta_\ell$, $\theta_K$, and $\phi$) and the invariant mass 
squared of the lepton pair ($q^2$). The definitions of the angles 
and the full angular distribution are described in Ref.~\cite{lhcbafb}.
For this fit we require $M_{\rm bc} >5.27~{\rm GeV}/c^2$, and 
the number of signal ($N_{\rm sig}$) and background events ($N_{\rm bkg}$)
are fixed to the values in Table~\ref{tab:kll_yields}.

\begin{figure}
\begin{center}
\includegraphics[width=0.55\textwidth]{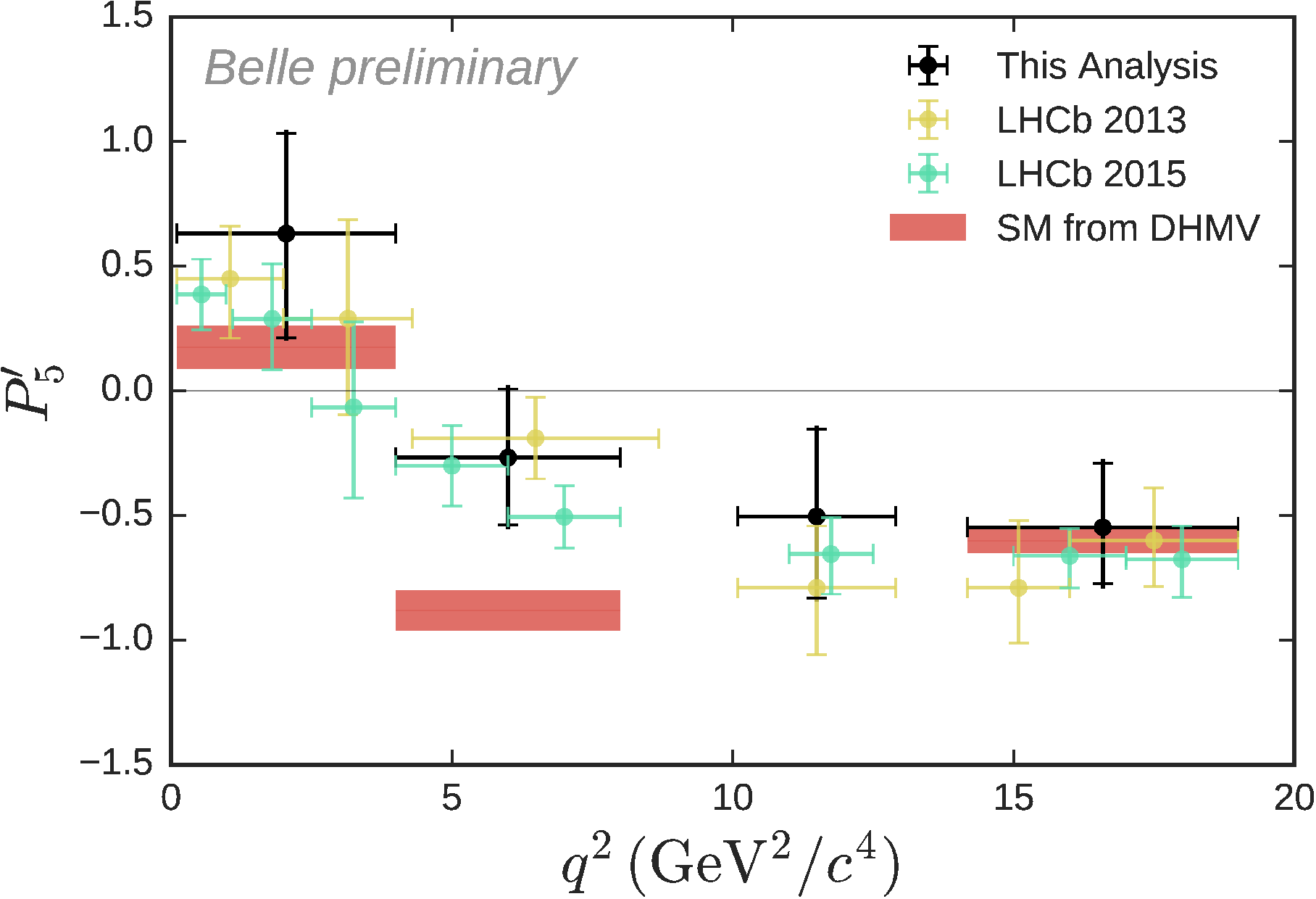}
\end{center}
\vskip-0.20in
\caption{\label{fig:kll_proj} 
Results for $P_5'$ as compared to SM predictions and to
previous measurements by LHCb~\cite{lhcb1}\cite{lhcb2}.}
\end{figure}

The angular observables $P_{i=4,5,6,8}'$ introduced in
Ref.~\cite{DescotesGenon1} contain information about 
short-distance effects. These observables are
considered to be largely free from form-factor 
uncertainties \cite{DescotesGenon2}, and thus 
they are a promising place to search for NP.
As the statistics in this analysis are insufficient 
to perform a full eight-dimensional fit, a ``folding technique''
is used as described in Ref.~\cite{lhcb1}.
We determine  $P_{4,5,6,8}'$ by performing a three-dimensional
unbinned maximum likelihood fit in four bins of $q^2$ 
using the folded signal probability density function and 
fixed background yields and shapes. We also fit for the 
longitudinal polarization $F_L$ and the transverse 
polarization asymmetry $A_T^{(2)}$. 
The fit results are shown in Fig.~\ref{fig:kll_proj}.
These results are consistent with SM predictions,
although one value of $P_5'$ differs by $2.1\sigma$ from 
the SM value. It is notable that this deviation is for the 
same $q^2$ bin and in the same  direction as a similar deviation 
reported by LHCb~\cite{lhcb1}\cite{lhcb2}.

\section{Measurement of the branching fraction of 
$\overline{B}{}^0 \rightarrow D^{*+} \tau^- \overline{\nu}_{\tau}$ 
relative to that of $\overline{B}{}^0 \ra D^{*+} \ell^- \overline{\nu}_\ell$}

Semi-tauonic $B$ decays of the type $b \rightarrow c \tau
\nu_{\tau}$ are sensitive probes to search for NP.
Charged Higgs bosons, which appear in supersymmetry and other 
models with at least two Higgs doublets, may induce measurable 
effects in the branching fraction due to the large mass of the 
$\tau$ lepton.  Similarly, leptoquarks, which carry both baryon 
number and lepton number, can also contribute to this process.  
The ratio of branching fractions
\begin{equation}
{\cal R}(D^{(*)}) = 
\frac{{\cal B}(\overline{B} \rightarrow D^{(*)} \tau^- \overline{\nu}_{\tau})}
{{\cal B}(\overline{B} \rightarrow D^{(*)} \ell^- \overline{\nu}_{\ell})} 
\hspace{0.8em}(\ell = e,\mu)
\label{eqn:RDstar}
\end{equation}
is typically measured instead of the absolute branching fraction 
${\cal B}(\overline{B}
\rightarrow D^{(*)} \tau^- \overline{\nu}_{\tau})$ to reduce 
systematic uncertainties arising from reconstruction efficiencies,
form factors, and the CKM matrix element $|V_{cb}|$. Standard Model
calculations predict ${\cal R}(D^*) = 0.252 \pm
0.003$~\cite{SM_PREDICTION_2} and ${\cal R}(D) = 0.297 \pm
0.017$~\cite{SM_PREDICTION_1}\cite{BELLE_HAD_NEW}.
Semi-tauonic $B$ decays were first observed by
Belle~\cite{BELLE_INCLUSIVE_OBSERVATION}, with subsequent
studies reported by Belle~\cite{BELLE_HAD_NEW}\cite{BELLE_INCLUSIVE},
\babar~\cite{BABAR_HAD_NEW}, and LHCb~\cite{LHCB_RESULT}. 
The world average values~\cite{HFAG} are
${\cal R}(D^*) = 0.322 \pm 0.018 \pm 0.012$ and 
${\cal R}(D) = 0.391 \pm 0.041 \pm 0.028$, which exceed 
the SM predictions by $3.0\sigma$ and $1.7\sigma$, respectively.  

Thus far, measurements of ${\cal R}(D^{(*)})$ at Belle 
and \babar\ have been performed using either a 
hadronic~\cite{BELLE_HAD_NEW}\cite{BABAR_HAD_NEW} or an 
inclusive~\cite{BELLE_INCLUSIVE_OBSERVATION}\cite{BELLE_INCLUSIVE} 
tagging method. Here we report the first measurement of ${\cal R}(D^*)$ 
using a semileptonic tagging method. We reconstruct signal events 
in modes in which one $B$ decays as
$\overline{B}{}^0 \rightarrow D^{*+} \ell^- \overline{\nu}_{\ell}$
and the the other $B$ decays as
$\overline{B}{}^0 \rightarrow D^{*+} \tau^- \overline{\nu}_{\tau}$,
$\tau^- \rightarrow \ell^- \overline{\nu}_{\ell} \nu_{\tau}$.
To reconstruct normalization events corresponding to the
denominator in Eq.~(\ref{eqn:RDstar}),
we require that both $B$ mesons decay to 
$D^{*+} \ell^- \overline{\nu}_{\ell}$.

Signal and normalization events are identified 
using a neural network. The dominant background 
contribution arises from events with falsely reconstructed 
$D^{(*)}$ mesons. To separate signal and normalization 
events from backgrounds, we use the energy variable 
$E_{\rm ECL}$, which is defined as the sum of the 
energies of neutral clusters detected in the 
electromagnetic calorimeter (ECL) that are not 
associated with reconstructed particles. 

We extract the signal and normalization yields
using a two-dimensional extended maximum-likelihood fit to the 
neural network output NN and $E_{\rm ECL}$. 
The resulting yields of signal and normalization events
are $231 \pm 23({\rm stat})$ and $2800 \pm 57({\rm stat})$, 
respectively. The ratio ${\cal R}(D^*)$ is
$0.302 \pm 0.030 \pm 0.011$, where the first and second
errors correspond to statistical and systematic uncertainties,
respectively. Our measurement is $1.6 \sigma$ higher than the
SM prediction. 

We investigate the compatibility of the data with a
Type II two-Higgs doublet model (2HDM)~\cite{Tanaka:1994ay}
and with $R_2$-type and $S_1$-type
leptoquark models~\cite{Sakaki:2013bfa}.
We find that the data allows for additional contributions
from 2HDM scalar operators and also vector operators, but 
contributions from a tensor operator with $0.34<C^{}_T<0.39$, 
an $R^{}_2$-type leptoquark model with $0.34<C^{}_T<0.38$, or
an $S^{}_1$-type leptoquark model with $0.22<C^{}_T<0.28$ are
disfavored.


\section{Observation of the decay $B^0_s \to K^0 \overline{K}{}^0$}
\label{sec:bskk}

The two-body decays $B^0_s \to hh'$, where
$h^{\scriptscriptstyle(}\kern-1pt{}'\kern-1pt{}^{\scriptscriptstyle)}$
is either a charged $\pi^\pm$ or $K^\pm$, have all been observed~\cite{PDG}.
However, the neutral modes $\pi^0\pi^0$, $\pi^0 K^0$, and $K^0\overline{K}{}^0$ 
have not. The decay $B^0_s\to K^0\overline{K}{}^0$ is of particular interest
because the branching fraction is predicted to be large: 
$(16-27)\times10^{-6}$~\cite{SM-branching}. The presence of
non-SM particles or couplings could measurably affect this
value~\cite{Chang:2013hba}. 
The current upper limit, 
$\mathcal{B}(B^0_s\to K^0\overline{K}{}^0)<6.6\times 10^{-5}$ 
at 90\% C.L., was set by Belle using 
only $23.6~{\rm fb^{-1}}$ of data recorded at the
$\Upsilon(5S)$~\cite{Peng:2010ze}. Here we 
update that result using the full data set 
of $121.4~{\rm fb^{-1}}$~\cite{Pal:2015ghq}.
At the $\Upsilon(5S)$, $B^0_s\overline{B}{}^0_s$ pairs
are produced in three decay channels: $B^0_s\overline{B}{}^0_s$, $B^{*0}_s
\overline{B}{}^0_s$ or $B^0_s\overline{B}{}^{*0}_s$, and 
$B^{*0}_0\overline{B}{}^{*0}_s$. 
The latter two channels dominate, with production fractions 
of $f_{B^{*0}_s\overline{B}{}^0_s}=(7.3\pm1.4)\%$ and
$f_{B^{*0}_s\overline{B}{}^{*0}_s}=(87.0\pm1.7)$\%~\cite{Esen:2012yz}.

Candidate $K^0$ mesons are reconstructed via the decay 
$K^0_S\to\pi^+\pi^-$ using a neural network technique. To suppress 
background arising from $e^+e^-\to q\bar{q}~(q=u,d,s,c)$ production, 
a second NN  is used. The output of the latter ($C_{\rm NN}$) ranges 
from $-1$ for background-like events to $+1$
for signal-like events. We make a loose requirement 
$C_{\rm NN}>-0.1$ and then translate $C_{\rm NN}$ to the
variable $C'_{\rm NN}$ as described in Section~\ref{sec:rho_gamma}.

\begin{figure}
\begin{center}
\includegraphics[width=0.32\textwidth]{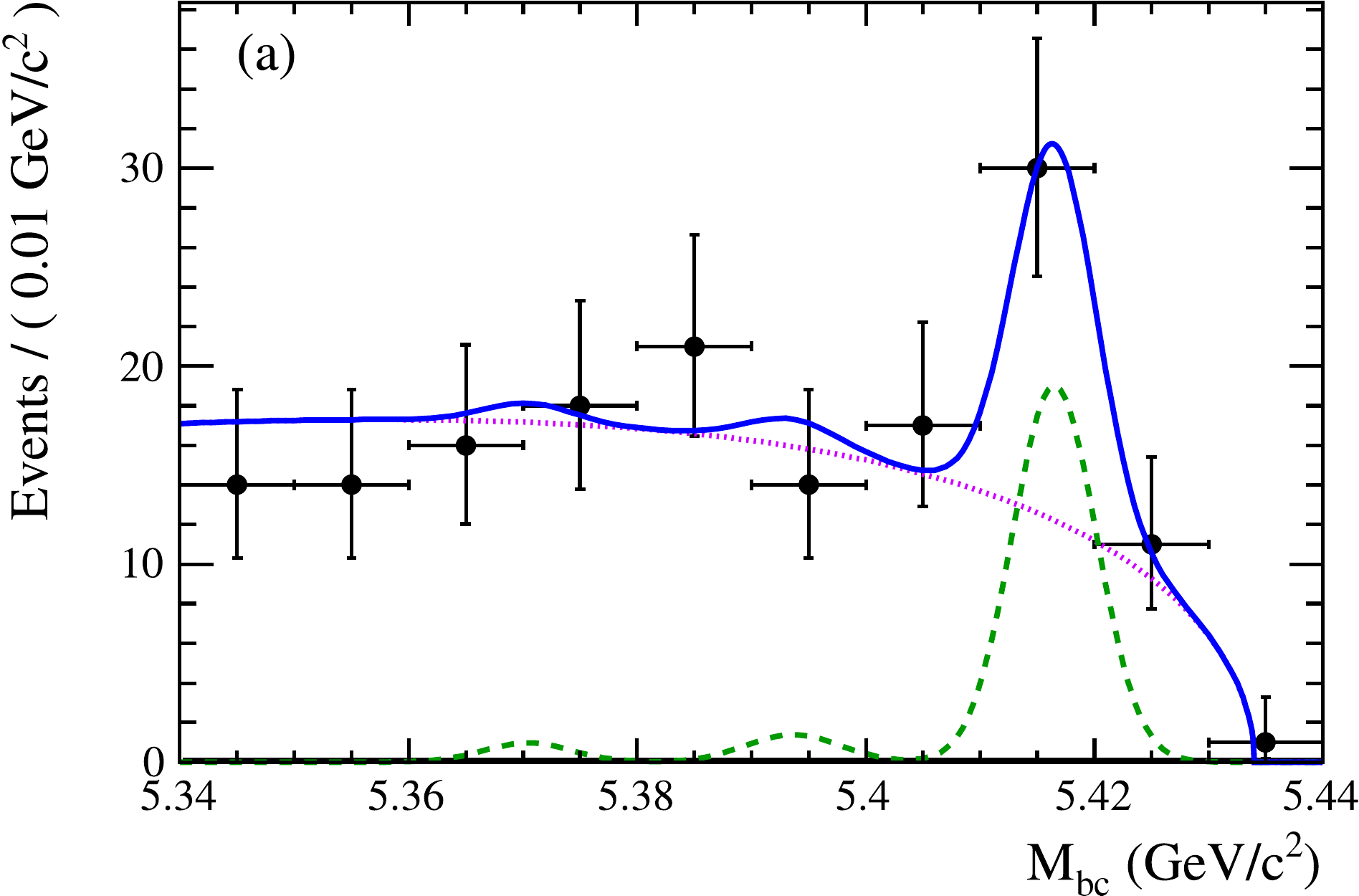}
\includegraphics[width=0.32\textwidth]{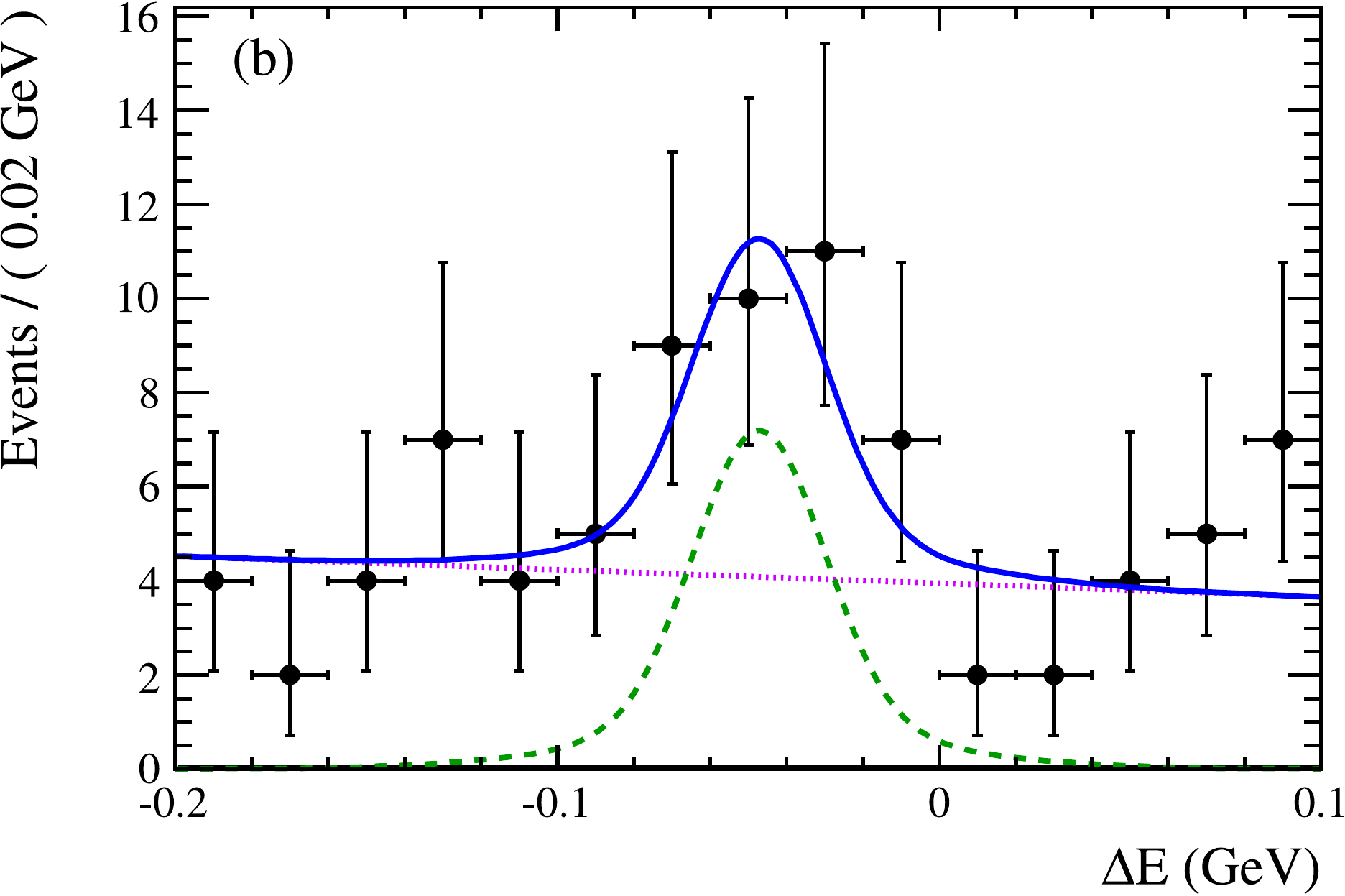}
\includegraphics[width=0.32\textwidth]{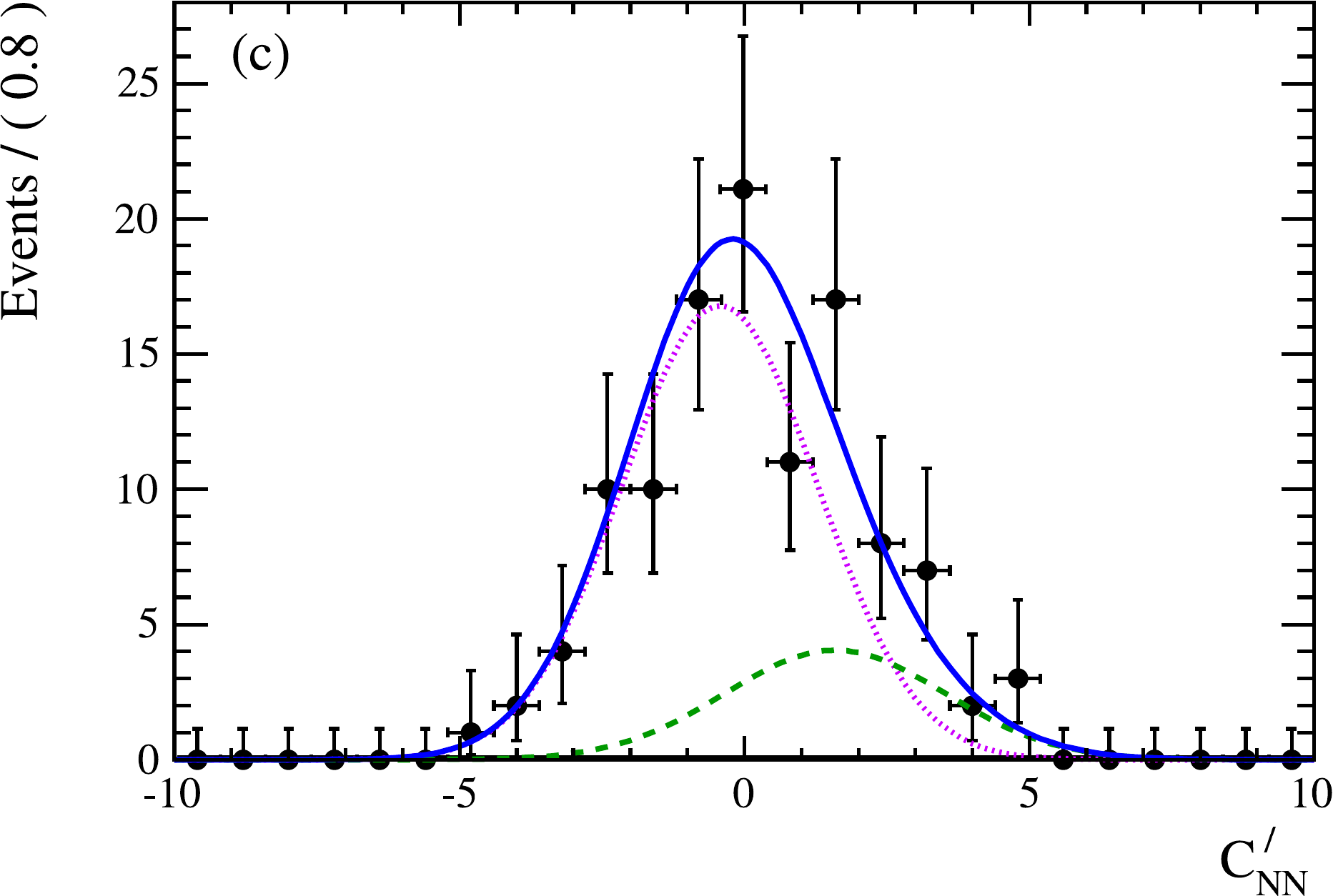}
\end{center}
\vskip-0.20in
\caption{\label{fig:bskk-fig2} 
Projections of the fit for $B^0_s \to K^0 \overline{K}{}^0$:
(a) $M_{\rm bc}$ for $-0.11~{\rm GeV} <\Delta E < 0.02~{\rm GeV}$ and
$C^{\prime}_{\rm NN}>0.5$; 
(b) $\Delta E$ for $5.405~{\rm GeV}/c^{2}<M_{\rm bc}< 5.427~{\rm GeV}/c^{2}$ 
and $C^{\prime}_{\rm NN}>0.5$; and
(c) $C^{\prime}_{\rm NN}$ for $5.405~{\rm GeV}/c^{2} < M_{\rm bc} < 
5.427~{\rm GeV}/c^{2}$ and $-0.11~{\rm GeV} <
\Delta E < 0.02~{\rm GeV}$. 
The points with error bars are data, the (green) dashed
curves show the signal, (magenta) dotted curves show the 
continuum background, and (blue) solid curves show the total. 
}
\end{figure}

We measure the signal yield by performing an unbinned 
extended maximum likelihood fit to variables $M_{\rm bc}$, 
$\Delta E$, and $C^{\prime}_{\rm NN}$. The results are
$29.0\,^{+8.5}_{-7.6}$ signal events and $1095.0\,^{+33.9}_{-33.4}$ 
background events. Projections of the fit result are shown in
Fig.~\ref{fig:bskk-fig2}.
The branching fraction is calculated as
\begin{eqnarray}
\mathcal{B}(B^0_s \to K^0 \overline{K}{}^0) & = &
\frac{N_{\rm sig}}{2 N_{B^0_s\overline{B}{}^0_s}(0.50)
\mathcal{B}^2_{K^0} \varepsilon}
\end{eqnarray}
where $N_{\rm sig}$ is the fitted signal yield; 
$N_{B^0_s\overline{B}{}^0_s}=(6.53\pm 0.66)\times10^6$ is the number 
of $B^0_s\overline{B}{}^0_s$ events~\cite{Oswald:2015dma};
$\mathcal{B}_{K^0}=(69.20\pm0.05)\%$ is the branching fraction for
$K^0_S\to\pi^+\pi^-$~\cite{PDG}; and $\varepsilon=(46.3\pm 0.1)\%$ is the
reconstruction efficiency as determined from MC simulation. The factor 0.50
accounts for the 50\% probability for $K^0\overline{K}{}^0\to K^0_S K^0_S$ (since
$K^0\overline{K}{}^0$ is $CP$ even). Inserting these values gives
$\mathcal{B}(B^0_s \to K^0 \overline{K}{}^0)=
(19.6\,^{+5.8}_{-5.1}\,\pm1.0\,\pm2.0)\times10^{-6}$, where the first
uncertainty is statistical, the second is systematic, and the third
reflects the uncertainty due to the total number of $B^0_s\overline{B}{}^0_s$
pairs. This result is in good agreement with the SM
prediction~\cite{SM-branching}.

The signal significance is calculated as
$\sqrt{-2\ln(\mathcal{L}_0/\mathcal{L}_{\rm max})}$,
where $\mathcal{L}_0$ is the likelihood value when the 
signal yield is fixed to zero, and $\mathcal{L_{\mathrm{max}}}$ 
is the likelihood value of the nominal fit. 
Systematic uncertainties are included in the significance by
convolving the likelihood function with a Gaussian function whose width
is equal to the systematic uncertainty associated with the signal 
yield. We obtain a signal significance of $5.1\sigma$, and thus 
our measurement constitutes the first observation of this decay.


\section{Measurement of the branching fraction and $CP$ asymmetry for $D^0 \ra V \gamma$}

$CP$ asymmetries in the charm sector are a promising area in which 
to search for new physics. The \cp\ asymmetry ($\mathcal{A}_{CP}$) of
many two-body $D$ decays has been measured~\cite{HFAG}, but all results
are consistent with zero. Here we present a first search for a \cp\ asymmetry 
in radiative $D^0 \ra V \gamma$ decays, where $V$ is a neutral vector 
meson $\phi$, $\overline{K}{}^{*0}\!$, or $\rho^0$.
Within the SM, such asymmetries are predicted to be small: 
${\cal A}^{}_{CP}\!\sim\!10^{-3}$.
However, theoretical studies~\cite{Isidori:2012yx}\cite{Lyon:2012fk} indicate
that extensions to the SM with chromomagnetic dipole operators can 
increase $\mathcal{A}_{CP}$ up to a few percent.
Several $D^0 \ra V \gamma$ decay modes have been measured 
by Belle and \babar~\cite{Abe:2003yv}\cite{Aubert:2008ai}. 
The current world average branching fractions are 
${\cal B}(D^0\ra\phi\gamma)=(2.70\pm0.35)\times 10^{-5}$ and 
${\cal B}(D^0\ra \overline{K}{}^{*0}\gamma)=(32.7\pm3.4)\times 10^{-5}$~\cite{PDG}.
The $D^0\ra\rho^0\gamma$ mode has not yet been observed, and
the current upper limit on the branching fraction is 
$\mathcal{B}(D^0 \ra \rho^0 \gamma) <24\times 10^{-5}$ 
at 90\%~C.L.~\cite{PDG}. 

Here we present new measurements of all three modes using the full 
Belle data set. Our results include the first measurement of the 
\cp\ asymmetries. For our analysis, the $D^0$ is required 
to originate from $D^{*+} \to D^0 \pi{}^+$ decays; this tags the 
$D^0$ flavor and suppresses combinatoric background. The vector 
mesons in the final state are reconstructed via $\phi \to K^+ K^-$,
$\overline{K}{}^{*0} \to K^-\pi^+$, and $\rho^0 \to \pi^+\pi^-$.
 
All branching fractions and $\mathcal{A}_{CP}$ are 
determined by normalizing to decay channels
$D^0\ra K^+K^-$ for $D^0\ra \phi\gamma$, 
$D^0\ra K^-\pi^+$ for $D^0\ra \overline{K}{}^{*0}\gamma$, 
and $D^0\ra \pi^+\pi^-$ for $D^0\ra \rho^0\gamma$.
The branching fractions are calculated as
\begin{equation}
\mathcal{B}_\mathrm{sig} =
\mathcal{B}_\mathrm{norm}\times 
\frac{N_\mathrm{sig}}{N_\mathrm{norm}}\times 
\frac{\varepsilon_\mathrm{norm}}{\varepsilon_\mathrm{sig}}
\end{equation}
where $\mathcal{B}$ and $N$ are the branching fraction
and fitted yield, respectively, of signal or normalization 
modes, and $\varepsilon$ is the reconstruction efficiency. 
For $\mathcal{B}_\mathrm{norm}$, the world-average 
value~\cite{PDG} is used. 

The extracted raw asymmetry 
\begin{equation}
 A_{\mathrm{raw}}=\frac{N(D^0) - N(\overline{D}{}^0)}{N(D^0)+N(\overline{D}{}^0)} 
 \end{equation}
has several contributions: 
$A_\mathrm{raw} = \mathcal{A}_{CP}+A_{\mathrm{FB}}+A^\pm_\varepsilon$,
where $A_\mathrm{FB}$ is the forward-backward production asymmetry
and $A^\pm_\varepsilon$ is the detection asymmetry between 
positively and negatively charged particles. Both of these 
asymmetries are eliminated by measuring $A^{}_{CP}$ of a
signal mode relative to that of its normalization mode, 
which has the same final-state charged particles.
The $CP$ asymmetry of a signal mode is then
$\mathcal{A}_{CP}^\mathrm{sig} = 
A^\mathrm{sig}_\mathrm{raw} - A^\mathrm{norm}_\mathrm{raw} + 
\mathcal{A}_{CP}^\mathrm{norm}$, 
where $\mathcal{A}_{CP}^\mathrm{norm}$ is the PDG value of
the $CP$ asymmetry for the normalization mode~\cite{PDG}.

To extract the signal yields and $CP$ asymmetries, we perform
a simultaneous two-dimensional unbinned extended maximum 
likelihood fit of the $D^0$ and $\overline{D}{}^0$ samples. 
The fit variables are the invariant mass of the $D^0$ and 
the cosine of the helicity angle $\theta_H$, defined as the angle 
between the $D^0$ and the positively or negatively charged hadron in 
the rest frame of the $V$ meson. For $D^0$ candidates we use the 
$K^+$/$K^-$/$\pi^+$ for $\phi$/$\overline{K}{}^{*0}$/$\rho^0$ decays, 
and we use the oppositely charged particles for $\overline{D}{}^0$ 
candidates. The fitted signal yields are 
$N^{}_{\phi\gamma} = 524\pm 35$, 
$N^{}_{K^{*0}\gamma} = 9104 \pm 396$, and
$N^{}_{\rho^0\gamma} = 500 \pm 85$.
The resulting branching fractions are
\begin{eqnarray*}
\mathcal{B}(D^0 \ra \phi \gamma) &= &(2.76 \pm 0.20 \pm 0.08) \times 10^{-5} \\
\mathcal{B}(D^0 \ra \overline{K}{}^{*0}\gamma) &= & 
(4.66 \pm 0.21 \pm 0.18) \times 10^{-4} \\
\mathcal{B}(D^0 \ra \rho^0 \gamma) & = & (1.77 \pm 0.30 \pm 0.08) \times 10^{-5} ,
\end{eqnarray*}
where the first uncertainty is statistical and the second is systematic. 
The result for $\phi\gamma$ is improved with respect to the 
previous result~\cite{Abe:2003yv}. The result for $\overline{K}{}^{*0}\gamma$ 
is 3.3$\sigma$ higher than the \babar\ result~\cite{Aubert:2008ai}. 
The result for $\rho^0\gamma$ is close to that for
$\phi\gamma$, which is consistent with theoretical expectations.
The significance is calculated as 
$\sqrt{-2\ln(\mathcal{L}_0/\mathcal{L_{\mathrm{max}}})}$
as described in Section~\ref{sec:bskk}.
Systematic uncertainties are included in the significance by 
convolving the likelihood function with a Gaussian whose width 
is equal to the systematic uncertainty associated with the signal 
yield. The significance for $\rho^0\gamma$ is $5.5\sigma$, and thus
our measurement constitutes the first observation of this decay.

The results for $\mathcal{A}_{CP}$ are
\begin{eqnarray*}
\mathcal{A}_{CP}(D^0 \ra \phi \gamma) & = & -0.094 \pm 0.066 \pm 0.001 \\
\mathcal{A}_{CP}(D^0 \ra \overline{K}{}^{*0} \gamma) & = & -0.003 \pm 0.020 \pm 0.000 \\
\mathcal{A}_{CP}(D^0 \ra \rho^0 \gamma) & = & \ \ 0.056 \pm 0.151 \pm 0.006 \,.
\end{eqnarray*}
No \cp\ asymmetry is seen in any of these radiative decays.

\section*{Acknowledgements}
The authors thank the BEACH 2016 organizers for a well-run workshop and
excellent hospitality. This research is supported by the U.S.\ Department 
of Energy.


\begin{thebibliography}{99}
\bibitem{charge-conjugates} 
Charge-conjugate modes are implicitly included unless noted otherwise
\end{thebibliography}
\end{document}